# A STATIC MALWARE DETECTION SYSTEM USING DATA MINING METHODS

Usukhbayar Baldangombo[1], Nyamjav Jambaljav[1], and Shi-Jinn Horng[2]

[1]Department of Communication Technology, School of Information Technology, National University of Mongolia
`{usukhbayar, nyamjav}@num.edu.mn`
[2]Department of Computer Science and Information Engineering, National Taiwan University of Science and Technology, Taipei, Taiwan
`horngsj@yahoo.com.tw`

*ABSTRACT*

*A serious threat today is malicious executables. It is designed to damage computer system and some of them spread over network without the knowledge of the owner using the system. Two approaches have been derived for it i.e. Signature Based Detection and Heuristic Based Detection. These approaches performed well against known malicious programs but cannot catch the new malicious programs. Different researchers have proposed methods using data mining and machine learning for detecting new malicious programs. The method based on data mining and machine learning has shown good results compared to other approaches.*

*This work presents a static malware detection system using data mining techniques such as Information Gain, Principal component analysis, and three classifiers: SVM, J48, and Naïve Bayes. For overcoming the lack of usual anti-virus products, we use methods of static analysis to extract valuable features of Windows PE file. We extract raw features of Windows executables which are PE header information, DLLs, and API functions inside each DLL of Windows PE file. Thereafter, Information Gain, calling frequencies of the raw features are calculated to select valuable subset features, and then Principal Component Analysis is used for dimensionality reduction of the selected features. By adopting the concepts of machine learning and data-mining, we construct a static malware detection system which has a detection rate of 99.6%.*

*KEYWORDS*

*Malware Detection, Malicious Codes, Malware, Malware Detection, Information Security, Data Mining*

## 1. INTRODUCTION

Cohen first formalized the term computer virus in 1983, it has evolved into malware detection [1]. Malicious programs, commonly termed as malware, can be classified into Virus, Trojans, Backdoors, and Logic bomb and a variety of other classes [2]. Computer malware remains a major thread to today's computer systems and networks. According to a recent threat report by Symantec, trends for 2011, created 403 million new malicious code signatures, which is a 41% increase over 2010, when 286 new malicious code signatures were added [3].

Although most common detection method is a signature based detection that makes the core of most commercial anti-virus programs, it is not effective against "zero-day attacks" or previously unknown malwares. The focus of malware researchers has shifted to find more generalized and scalable features that can identify previously unknown malwares instead of a static signature. Two basic types of approaches are used to analyse malware that are static and dynamic analysis. In the static analysis code and structure of a program is examined without actually running the program while dynamic analysis the program is executed in a real or virtual environment.

In this work we propose a static malware detection system using data mining techniques that it automatically extracts features of malware and clean programs and uses them to classify programs into malicious or benign executables. Our goal in the evaluation of this system is to simulate the task of detecting new malicious executables.

## 2. RELATED WORKS

We consider details of most relevant malware detection techniques in this section. In recent years many malware researchers have focused on data mining to detect unknown malwares. Data mining is the process of analyzing electronically stored data by automatically searching for patterns [5]. Machine learning algorithms have been used widely for different data mining problems to detect patterns and to find correlations between data instances and attributes. Many researchers have used n-grams or API calls as their primary type of feature that are used to represent malware instances in a suitable format for data mining purposes.

Shultz et al. [6] proposed a method using data mining techniques for detecting new malicious executables. Three different types of features are extracted from the executables, i.e. the list of DLLs used by the binary, the list of DLL function calls, and number of different system calls used within each DLL. Also they analyze byte sequences extracted from the hexdump of an executable. The data set consisted of 4,266 files out of which 3,265 were malicious and 1,001 were legitimate or benign programs. A rule induction algorithm called Ripper [7] was applied to find patterns in the DLL data. A learning algorithm Naïve Bayes (NB), which is based on Bayesian statistics, was used to find patterns in the string data and n-grams of byte sequences were used as input data for the Multinomial Naïve Bayes algorithm. A data set is partitioned in two data sets, i.e., a test data set and a training data set. This is to allow for performance testing on data that are independent from the data used to generate the classifiers. The Naïve Bayes algorithm, using strings as input data, yielded the highest classification performance with an accuracy of 97.11%. The authors compared their results with traditional signature-based methods and claimed that the data mining-based detection rate of new malware was twice as high in comparison to the signature-based algorithm.

A similar approach was used by J. Z. Kolter et al. [8], where they use n-gram analysis and data mining approaches to detect malicious executables in the wild. The authors used a hexdump utility to convert each executable to hexadecimal code in an ASCII format and produced n-gram features by combining each four-byte sequence into a single term. Their primary dataset consisted of 1971 clean and 1651 malicious programs They used different classifiers including Instance-based Learner, TFIDF, Naive-Bayes, Support vector machines, Decision tree, boosted Naive-Bayes, SVMs and boosted decision tree. They used information gain to select valued features which are provided as input to all classifiers. The area under an ROC curve (AUC) is a more complete measure compared with the detection accuracy as they reported [9]. AUCs show that the boosted decision trees outperform rest of the classifiers for both classification problems.

M. Siddiqui et al. [10] used Data Mining for detection of Worms. They used variable length instruction sequence. Their Primary data set consists of 2,775 Windows PE files, in which in which 1,444 were worms and 1,330 were benign. They performed detection of compilers, common packers and crypto before disassembly of files. Sequence reduction was performed and 97% of the sequences were removed. They used Decision Tree, Bagging and Random Forest models using. Random forest performed slightly better than the others.

## 3. MALWARE DETECTION SYSTEM'S ARCHITECTURE

We describe the architecture of our proposed system in this section. This system is to accurately detect new malware (unknown malware) binaries using a number of data mining techniques.

Figure 3 shows the architecture of our malware detection system. The system consists of three main modules: (1) PE-Miner, (2) feature selection and data transformation, and (3) learning algorithms.

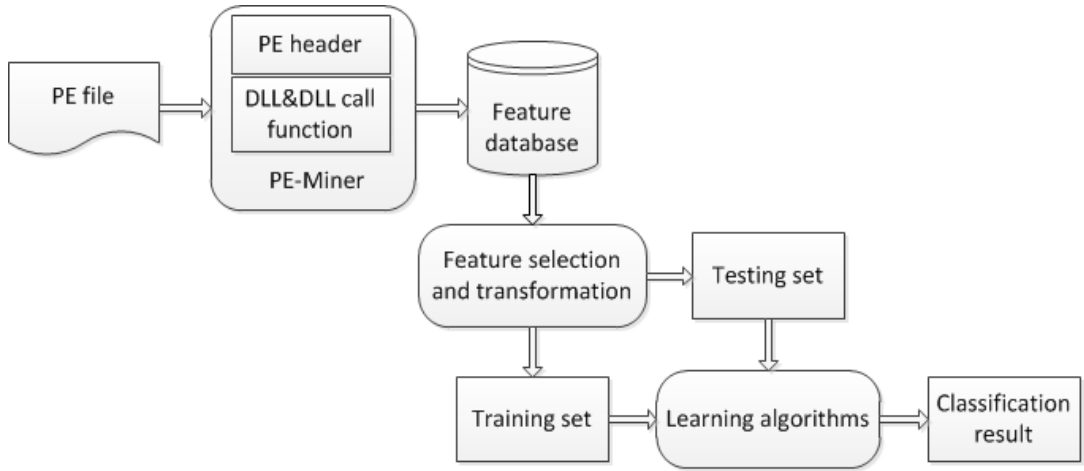

Figure 1. Architecture of the Malware Detection System

**Module 1:** PE-Miner parses PE tables of all executables to find out all PE header information, DLL names, and API functions inside each DLL as raw features;

**Module 2:** Information Gain value of each PE header, and calling frequency of each DLL and API function inside DLL are computed and those raw features with Information Gain values and calling frequencies greater than a threshold are selected; then Principal Component Analysis is applied to further transform and reduce the number of features;

**Module 3:** According to the features after PCA, transform every program in the database to its corresponding feature vector and then use a learning algorithm to derive a classification result from these labeled feature vectors;

### 3.1. Dataset description

Our dataset consists of a total of a total 247348 Windows programs in PE file format that are 236756 malicious and 10592 clean programs. There were no duplicate programs in our dataset. The malicious programs were obtained from several sources such as VX Heavens Virus Collection database, which is available for free download in the public domain [4], our local laboratory, and our colleague's laboratory in Taiwan. The numbers of malicious programs are grouped by their corresponding family name in Table 1. The clean programs were collected from Download.com [11], Windows' system files and our local laboratory's program files.

We only consider non-packed Windows binary files in PE format. Some malware writers compress their code to make them undetectable by antivirus programs. So we first used number of most popular tools to decompress the files, which are compressed, such as UPX, ASPack, PECompact, etc. that make the PE files recognizable by PE parser. We developed a PE file parser named PE Miner, which extracts all PE format file's header information, DLLs, and API functions inside each DLL of the programs in the dataset.

Table 1. Number of malicious programs grouped by their family name

| Name of malicious program | Counts | Name of malicious program | Counts |
|---|---|---|---|
| Backdoor | 51,457 | Trojan-IM | 481 |
| Trojan | 46,142 | Packed | 379 |
| Trojan-Downloader | 44,410 | IM-Worm | 379 |
| Trojan-GameThief | 29,050 | not-virus_Hoax | 318 |
| Virus | 26,941 | Trojan-Mailfinder | 301 |
| Trojan-PSW | 16,889 | Flooder | 267 |
| Trojan-Spy | 11,748 | DoS | 257 |
| Trojan-Dropper | 8,193 | Email-Flooder | 196 |
| Trojan-Banker | 6,986 | Trojan-DDoS | 135 |
| Worm | 6,152 | IM-Flooder | 134 |
| Email-Worm | 3,314 | Trojan-Ransom | 94 |
| Rootkit | 3,210 | SpamTool | 64 |
| Trojan-Clicker | 2844 | Trojan-Notifier | 59 |
| Trojan-Proxy | 2045 | SMS-Flooder | 57 |
| Exploit | 1938 | Spoofer | 26 |
| Net-Worm | 1938 | not-virus_BadJoke | 20 |
| Hoax | 1216 | Trojan-SMS | 14 |
| Constructor | 933 | Unknown | 2 |
| HackTool | 775 | Sniffer | 2 |
| VirTool | 604 | not-a-virus_NetTool | 1 |
| P2P-Worm | 573 | not-a-virus_RemoteAdmin | 1 |
| IRC-Worm | 545 | | |

### 3.2 Feature extraction

The information of an executable can be easily derived its PE file format without running programs [12,13]. We developed an application named "PE-Miner" to parse PE tables of Windows executable programs in the collected dataset. The PE Miner extracts all PE header information, DLL names, and API function names called within those DLLs contained in a PE file. The PE-Miner program stores all the extracted information in a database. In this study, the PE-Miner program extracted all raw features of 236756 malicious and 10592 clean programs and stored them in a database. These features are summarized in Table 2.

Table 2. List of raw features extracted from PE files

| № | Description | Data Type | Quantity |
|---|---|---|---|
| 1 | Unique DLL features | Binary | 792 |
| 2 | Unique API function calls | Binary | 24,662 |
| 3 | PE DOS header | Integer | 31 |
| 4 | PE file header | Integer | 7 |
| 5 | PE optional headers | Integer | 30 |
| 6 | PE data directories | Integer | 16 |

| 7 | PE section headers | Integer | 10 |
| 8 | PE import description | Integer | 5 |
| 9 | PE export description | Integer | 11 |
| 10 | PE resource table | Integer | 6 |
| 11 | PE debug info | Integer | 8 |
| 12 | PE delay import | Integer | 8 |
| 13 | PE TLS table | Integer | 6 |
| | **Total** | | **25,592** |

Figure 2 shows main graphic user interface (GUI) of the PE-Miner that extracts raw features from PE files in a specified folder and automatically stores all features in a database.

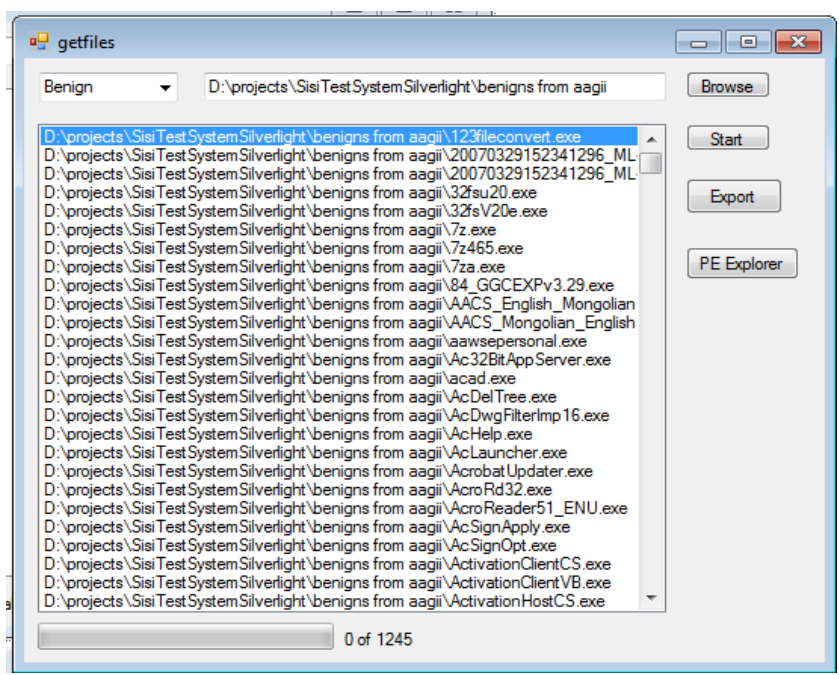

Figure 2. PE-Miner program's main GUI

### 3.3 Feature selection

The data must be pre-processed to select a subset of the attributes to use in learning because many practical situations have far too many attributes in for learning schemes to handle, and some of them-perhaps the overwhelming majority-are clearly irrelevant or redundant. In practice learning methods performance can frequently be improved by preselecting, however learning methods themselves try to select attributes appropriately and ignore irrelevant or redundant ones, [14].

After raw features of all executable programs in the database have been extracted using PE-Miner program, the system computes each PE header feature's Information Gain value by Equation (3) in Subsection 3.1, and calling frequency of DLLs and API functions. The top-20 PE header features ranked by IG on dataset are presented in Table 3.

Table 3. List of the top ranked 20 features of the PE header

| № | Features | Mean values | |
| | | Malware | Benign |
| --- | --- | --- | --- |
| 1 | SizeOfInitializedData | 59,675,152 | 600,081 |

| 2  | NumberOfSymbols         | 59,251,515  | 1,655,386 |
| 3  | certificateTableSize    | 103,615,710 | 388,027   |
| 4  | exporttableSize         | 8,350,672   | 1,702     |
| 5  | IATSIZE                 | 2,341,313   | 509       |
| 6  | BoundImportSize         | 68,707,156  | 200,506   |
| 7  | LoadConfigTableSize     | 86,574,977  | 306,799   |
| 8  | BaseRelocationTableSize | 58,377,700  | 228,904   |
| 9  | e_cp                    | 1,273       | 5         |
| 10 | CLRRuntimeHeaderSize    | 49,875,208  | 135,841   |
| 11 | VirtualSize             | 60,309,693  | 301,690   |
| 12 | SizeOfRawData           | 80,791,328  | 1,817,641 |
| 13 | NumberOfLineNumbers     | 1,675       | 3         |
| 14 | e_cblp                  | 1,353       | 143       |
| 15 | e_minalloc              | 1,118       | 2         |
| 16 | NumberOfRelocations     | 1,167       | 67        |
| 17 | AddressOfFuncitons      | 2,884,543   | 378,963   |
| 18 | AddressOfNameOrdinals   | 2,538,096   | 382,150   |
| 19 | AddressOfNames          | 2,624,183   | 11,265    |
| 20 | e_ovno                  | 1,409       | 10        |

As shown in Table 3, there is a significant difference the values of major linker versions and the size of the initialized data in the benign and malicious executables. The size of the initialized data in benign executables is usually significantly higher compared to those of the malicious executables. It is interesting to note that a reasonable number of symbols are present in benign executables. It can be seen in Table 3 that the values of fields such as number of relocation and a number of line numbers are significantly higher in the malware executable in comparison. Note that the number of functions is relatively higher for the malware executables.

Considering every DLL as a feature seems to be useful because a program's DLLs can tell you a lot about its functionality. The top-30 DLL name features ranked by calling frequency on dataset are presented in Table 4. Most commonly used DLL is Kernel32.dll that contains core functionality, such as access and manipulation of memory, files, and hardware.

Table 4. List of the top ranked 30 DLL names by calling frequency

| No  | DLLs         | Frequency | Description                                         |
|-----|--------------|-----------|-----------------------------------------------------|
| 1.  | kernel32.dll | 3,467,435 | Windows NT BASE API Client DLL                      |
| 2.  | user32.dll   | 912,115   | USER API Client DLL                                 |
| 3.  | advapi32.dll | 441,914   | Advanced Win32 application programming interfaces   |
| 4.  | gdi32.dll    | 201,687   | GDI Client DLL, involved in graphical displays.     |
| 5.  | wininet.dll  | 84,100    | Internet Extensions for Win32                       |
| 6.  | comctl32.dll | 72,010    | User Experience Controls Library                    |
| 7.  | shell32.dll  | 55,169    | Windows Shell Common DLL                            |
| 8.  | wsock32.dll  | 49,133    | Windows Socket 32-Bit DLL                           |
| 9.  | oleaut32.dll | 49,060    | OLE 2 - 32 automation                               |
| 10. | msvbvm50.dll | 44,851    | Visual Basic Virtual Machine                        |
| 11. | ole32.dll    | 43,531    | 32-bit OLE 2.0 component                            |
| 12. | shlwapi.dll  | 38,718    | Library for UNC and URL Paths, Registry Entries     |
| 13. | ws2_32.dll   | 22,486    | Windows Socket 2.0 32-Bit DLL                       |
| 14. | ntdll.dll    | 18,570    | Win32 NTDLL core component                          |
| 15. | urlmon.dll   | 16,117    | OLE32 Extensions for Win32                          |

| 16. | version.dll | 12,481 | Version Checking and File Installation Libraries |
| --- | --- | --- | --- |
| 17. | crtdll.dll | 11,420 | Microsoft C Runtime Library |
| 18. | comdlg32.dll | 10,484 | Common Dialogue Library |
| 19. | winnm.dll | 8,793 | Windows Multimedia API |
| 20. | rpcrt4.dll | 5,531 | Remote Procedure Call Runtime |
| 21. | psapi.dll | 5,228 | The process status application programming interface |
| 22. | msvcr100.dll | 4,702 | Microsoft Visual C++ Redistributable DLL |
| 23. | hal.dll | 3,993 | The Windows Hardware Abstraction Layer |
| 24. | mpr.dll | 3,670 | Multiple Provider Router DLL |
| 25. | netapi32.dll | 3,510 | Net Win32 API DLL |
| 26. | avicap32.dll | 3,177 | AV1 capture API |
| 27. | rasapi32.dll | 2,973 | Remote Access 16-bit API Library |
| 28. | cygwin1.dll | 2,803 | Cygwin® POSIX Emulation DLL |
| 29. | mscoree.dll | 2,774 | Microsoft .NET Runtime Execution Engine |
| 30. | imagehlp.dll | 2,049 | Windows NT Image Helper |

Using Windows' advanced core components such as the Service Manager and Registry are provided by Advapi32.dll. User32.dll contains all the user-interface components, such as buttons, scroll bars, and components for controlling and responding to user actions. Gdi32.dll contains functions for displaying and manipulating graphics. Executables generally do not import Ntdll.dll directly, although it is always imported indirectly by Kernel32.dll. It is a file containing the interface to the Windows kernel. If an executable imports this file, it means that the author intended to use functionality not normally available to Windows programs. Some tasks, such as hiding functionality or manipulating processes, will use this interface. WSock32.dll and Ws2_32.dll are networking DLLs. Programs connect to a network or perform network-related tasks that access either of these files. Higher-level networking functions that implement protocols such as FTP, HTTP, and NTP are contained in Wininet.dll. The last domain of relevant features is API call categories (see in Figure 3).

kernel32.dll.LoadLibraryA;kernel32.dll.GetProcAddress;….

advapi32.dll.RegSetValueExA;advapi32.dll.RegOpenKeyExBookmark;…

user32.dll.MessageBoxA; user32.dll.GetKeyboardType; …

wsock32.dll.recv; wsock32.dll.send; …

Figure 3. API functions called inside each DLL

As a result of our experimental analysis on the malicious executables in the dataset, we have identified seven main groups of commonly used API function call features that are based on the malicious behaviours and these are listed below:

**File-related behaviour:** Create a specific file in sensitive folders; Delete, break, cover system files or application files; Edit file; Traverse file directory and search for target files;

**Process-related behaviour:** Release DLL file to inject into system process and create a new thread to hide itself; Create new processes to execute its code; Create threads to search and terminate the process of anti-virus or protection software; Create a matrix process to prevent repeated execution;

**Memory-related behaviour:** Free, move and replace spaces of the memory; Occupy extra memory space, decrease the total available memory size; Forbid memory allocation and reclaim memory space; Point the interrupt vector address to the initial address of the malicious code;

**Register-related behaviour:** Add or delete system service; Auto run while the system is starting up; Hide and protect itself; Undermine system function;

**Network-related behaviour:** Open or listen specific port; Send through chatting software; Steal information and sending them to the mailbox; Enumerate weak password vulnerable computers; Occupying resources;

**Windows service behaviour:** Terminating windows update service; Terminating windows firewall; Opening Telnet service;

**Other behaviours:** Hooking keyboard; Hiding window; Alter system time to disable software prevention; Restart the computer; Scan existing vulnerabilities of the system;

To interpret these features, when going back to malware definitions and their known major behaviours we can conclude that file management is the most important activity in viruses, because this kind of malware is known to copy itself multiple times and mutate itself every now and then if it's a polymorphic/metamorphic malware. Another difference when studying malware and benign applications is the type of process management done by malicious code that is intended to hide from, or spoof antivirus tools and process monitors. Since getting hidden is more complicated than running normally, this approach needs more calls to process management procedures. The third discriminative category is the Memory, which shows the difference that malicious executable intends to free, move and replace spaces of the memory. Next category is the Registry, malicious executables have heavily used the Windows Registry to change system menus, user privileges, startup programs, and file extension handling attributes to ease malware distribution or to destruct the target computer and so on next category behaviours are activated.

According to PE header, DLL, and API function's feature sets constructed above, we could use a feature vector to represent each program Pi. Each vector element corresponds to a feature $F_j$. And its value indicates whether that executable has the corresponding feature or not. With these values, the feature vector F for a program P is defined as follows:

$$F = \{F_1, F_2, F_3, \ldots F_n\}$$

## 4. EXPERIMENTAL RESULTS

Our dataset consists of a total of a total 247348 Windows programs in the PE file format that are 236756 malicious and 10592 clean programs. There were no duplicate programs in our dataset. The malicious programs were obtained from several sources such as VX Heavens Virus Collection database, our local laboratory, and our colleague's laboratory in Taiwan. The numbers of malicious programs are grouped by their corresponding family name in Table 1. The clean programs were collected from Download.com, freshly installed Windows's system files and local laboratories. This is a comprehensive database that contains many thousands of malware samples. The samples consist of Virus, Worm, Trojan, Backdoors, and Spyware etc..

We only consider non-packed Windows binary files in PE format. Some malware writers compress their code to make them undetectable by antivirus programs. So we first used a number of the most popular tools to decompress the files, which are compressed, such as UPX, ASPack, PECompact, etc. that make the PEs recognizable by the PE parser. We have selected to evaluate the performance of the following classifiers: SVM, J48, and NB rule learner. All of these classifiers are available as part of the Java-based open source machine learning toolkit Weka [14]. We have done our experiments on an Intel Pentium Core 2 Duo 2.33 GHz processor with 2GB RAM. The Microsoft Windows 7 SP1 is installed on this machine.

We trained and tested the classifiers (SVM, J48, NB), with cross-validation 10 folds, i.e., the dataset is randomly divided into 10 smaller subsets, where 9 subsets are used for training and 1 subset is used for testing. The process is repeated 10 times for every combination. This methodology helps in evaluating the robustness of a given approach to detect malicious Win32 files that contain malware without any a priori information.

### 4.1 Performance evaluation criteria

We have used the standard 10-fold cross-validation process in our experiments: the dataset is randomly divided into 10 smaller subsets, where 9 subsets are used for training and 1 subset is used for testing. The process is repeated 10 times for every combination. To evaluate our system we were interested in several quantities listed below:

**True Positives (TP):** the number of malicious executable examples classified as malicious executable

**True Negatives (TN):** the number of benign programs classified as benign

**False Positives (FP):** the number of benign programs classified as malicious executable

**False Negatives (FN):** the number of malicious executable classified as benign executable

We were interested in the detection rate of the classifiers. In our case this was the percentage of the total malicious programs labelled malicious. We were also interested in the false positive rate. This was the percentage of benign programs which were labelled as malicious, also called false alarms.

True positive rate is also called the Detection Rate (DR) and defined as

$$\frac{TP}{TP+FN} * 100\%,$$

False Positive Rate (FPR) as

$$\frac{FP}{TN+FP} * 100\%,$$

and Overall Accuracy (OA)

as $$\frac{TP+TN}{TP+TN+FP+FN} * 100\%.$$

Evaluating the detection efficiency of the system, it is not only considering the higher TP rate but also considering the lower FP rate. Therefore, the overall accuracy could be the main reference for us to compromise both TP rate and FP rate.

### 4.2 Results

There were total 138 PE header, 792 DLL, and 24662 API function's raw features in our feature set. We did some experiment to select a subset of features by the detection rate of the system. We just selected a number of features with different ranges, trained the system, and tested it to see the overall accuracy. According to the result of our analysis, we finally selected the top 88 PE header, 130 DLL, and 2453 API function features to train our system which had better performances with the classifiers. A number of selected subset features of the experiments are given in Table 5. A list of the full classification results are shown in Table 7.

Table 5. J48 classifier's experimental results of the subset feature selection

| Feature name | # (features) | DR (%) | FPR (%) | OA (%) |
|---|---|---|---|---|
| PE header | 88 | 99.1 | 2.7 | 97.8 |
| DLLs | 130 | 96.2 | 8.6 | 95.4 |
| API functions | 2,453 | 97.8 | 5.3 | 96.5 |

In order to know the contribution of PCA transformation, we first removed PCA function from the system, i.e., the selected features were directly applied to train the classifiers. When we enabled PCA function, we got better results as shown in Table 6. The training time was much lessened from a couple of days to several hours because of the number of features being transformed and reduced, and the performances were raised as well.

TABLE 6. The effect of PCA upon the performance of J48 classifier

| Feature name | # (feature) | | DR (%) | |
|---|---|---|---|---|
| | Before PCA | After PCA | Before PCA | After PCA |
| PE header | 88 | 39 | 99.1 | 99.5 |
| DLLs | 130 | 75 | 96.2 | 97.5 |
| API functions | 2,453 | 307 | 98.9 | 99.3 |

Finally, we have three different kinds of features from the Windows executables in our dataset. We combined these features with each other into one feature set, which we call the hybrid feature set (HFS), that we are interested in knowing if we use a hybrid feature set, whether the detection rate of the system is improved or not. Table 7 shows the complete classification results of the system.

Table 7. List of the system performance by individual and combined features

| Feature type | Classifier | DR (%) | FPR (%) | OA (%) |
|---|---|---|---|---|
| PE header | Naïve-Bayes | 96 | 9.1 | 77 |
| | SVM | 97 | 7.6 | 80 |
| | J48 | 99.5 | 2.7 | 99 |
| API functions | Naïve-Bayes | 88 | 7.7 | 97 |
| | SVM | 97 | 7.1 | 81 |
| | J48 | 99.3 | 5.3 | 99.1 |
| DLLs | Naïve-Bayes | 65 | 8.0 | 89 |
| | SVM | 70 | 5.6 | 91 |
| | J48 | 97.5 | 8.6 | 96.9 |
| **Hybrid feature** | | | | |
| PE header&DLLs | Naïve-Bayes | 95 | 8.9 | 72 |
| | SVM | 96 | 7.3 | 85 |
| | J48 | 97.1 | 4.7 | 90 |
| PE header&API functions | Naïve-Bayes | 94 | 8.2 | 93 |
| | SVM | 98 | 5.3 | 97 |

|       | J48         | 99.6 | 2.7  | 99 |
|-------|-------------|------|------|----|
|       | Naïve-Bayes | 94   | 9. 2 | 93 |
| ALL   | SVM         | 97   | 4. 5 | 97 |
|       | J48         | 97.6 | 2.8  | 98 |

As shown in Table 7 the detection accuracy of J48 classifier outperforms the rest of the data mining classifiers in most of the cases. We can conclude from Table 7 that the classifiers using the DLL name feature had the lowest detection rate. Further, Naïve-Bayes gives the worst detection accuracy in most cases. The classifiers using combined PE header and API function features give slightly better detection accuracy than the classifiers using PE header feature alone. As the result we can conclude that the PE header feature without combining with the other features can be effectively used for detecting zero day malwares. Moreover, a classifier using PE header feature can have the smallest processing overheads.

**4.3 Implementation**

We implement an application program named malDetection which is integrated several approaches that we have mentioned in the previous sections. Therefore, first, we export a model of trained J48 classifier from the WEKA. And then we integrate the model and the other functions, which extract, select, and reduce features of PE file, into an application program. Figure 4 shows main GUI of the malDetection program.

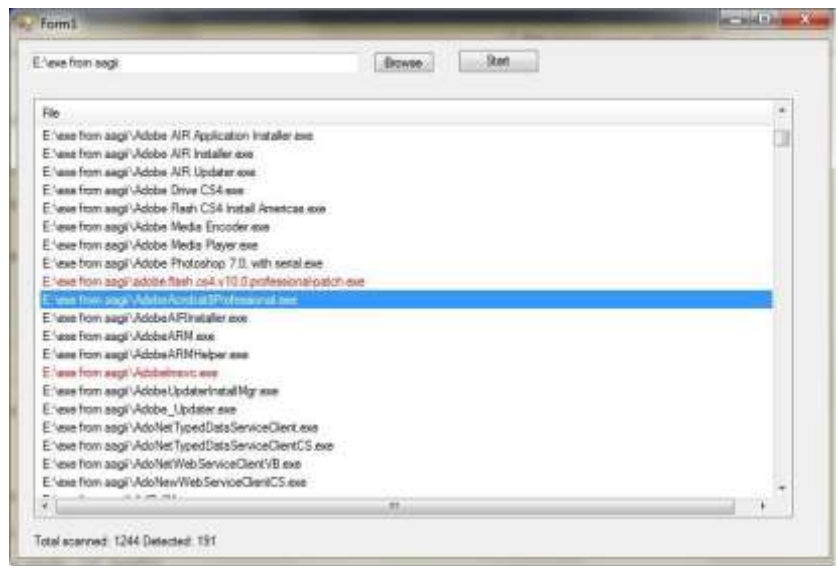

Figure 4. Main GUI of the malDetection program

The malDetection program's working steps are illustrated in Figure 5.

**Step 1:** It browses specified directory and its subdirectories to find files in the directory; if a file in the directory is not PE format file, it just ignores the file and continues its work;

**Step 2:** It parses PE tables of the executable to find out all PE header information, DLL names, and API functions inside each DLL as raw features;

**Step 3:** It selects the specified features from the raw features;

Step 4: Principal Component Analysis is applied to further transform and reduce the number of features;

**Step 5:** It transforms a program's data to its corresponding feature vector; and then apply to a classification algorithm;

**Step 6:** It prints the classification results as shown at the bottom of Figure 4;

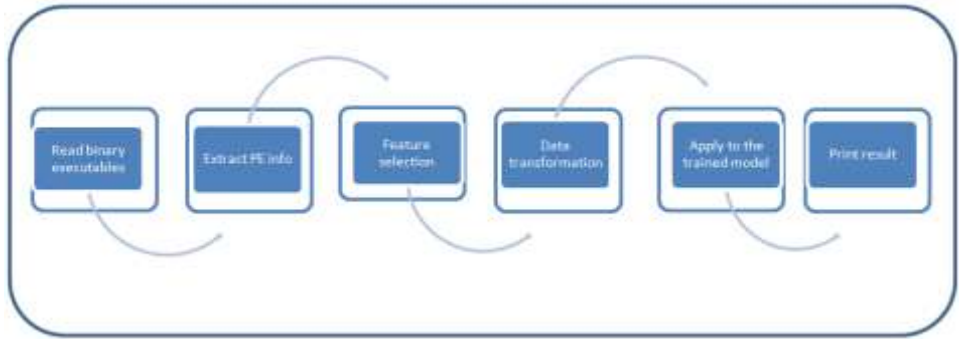

Figure 5. Working steps of the malDetection program

Finally, we prepared a standalone setup project from the malDetection program' source file.

## 5. CONCLUSIONS

This work presented a static malware detection system using data mining techniques. This system is based on mining PE header information, DLL name, API function calls inside each DLL features of the executables for detecting malware and malicious codes. We developed a PE-Miner program to parse PE format of the Windows executables in our dataset. The PE Miner extracts all PE header information, DLL names, and API function calls inside each DLL contained in a PE file. And this program stores all the extracted information in a database.

After raw features of all executable programs in the database have been extracted using PE-Miner program, the system computes each PE header feature's Information Gain values and counts the calling frequencies of DLLs and API functions to select subset feature sets. According to the result of our analysis, we finally take top 88 PE header, 130 DLL, and 2453 API function features to train our system which had better performances with the classifiers.

In order to know the contribution of PCA transformation, we first removed PCA function from the system, i.e., the selected features were directly applied to train the classifiers. When we enabled PCA function, we got better results and the training time was much lessened from a couple of days to several hours because of the number of features being transformed and reduced, and the performances were raised as well. Finally, we combine the three different kinds of features into one feature set, which we call the hybrid feature set (HFS), and these features are used to train SVM, J48, and NB classifiers respectively for detection of Windows PE malwares.

As shown the experimental results, the detection accuracy of J48 classifier outperforms the rest of the data mining classifiers in most of the cases. And, the classifiers using the DLL name feature had the lowest detection rate. Further, Naïve-Bayes gives the worst detection accuracy in most cases. The classifiers using combined PE header and API function features give slightly better detection accuracy than the classifiers using PE header feature alone. As the result we can conclude that the PE header feature without combining with the other features can be effectively used for detecting zero day malwares. Moreover, a classifier using PE header feature can have the smallest processing overheads. The experimental results show that our system's the best detection accuracy is 99.6% with the false positive rate 2.7%.